\documentstyle[preprint,aps]{revtex}
\begin{document}
\draft
\title{
Competition between plaquette and dimer phases\\
 in Heisenberg
chains}

\author{N. B. Ivanov}
\address{Institute  for Solid  State
Physics, Bulgarian Academy of Sciences,\\
Tzarigradsko
chaussee-72, 1784 Sofia, Bulgaria}

\author{J. Richter}
\address{Institut f\"ur Theoretische Physik,
Universit\"at Magdeburg,\\
 P.O.Box 4120, D-39016 Magdeburg, Germany}
\date{\today}
\maketitle
\begin{abstract}
We consider a class of one-dimensional Heisenberg
spin models (plaquette chains) related to the recently found
$1/5$-depleted
square-lattice Heisenberg system $CaV_4O_9$. A number of exact and
exact-numerical results concerning the properties of the
competing dimer and resonating plaquette phases
are presented.

{\bf Author for correspondence}: N. B. Ivanov\\
Institute of Solid State Physics,
Bulgarian Academy of Sciences,\\
72 Tzarigradsko chaussee blvd.
1784 Sofia, Bulgaria, \\
fax:00359 2 9753 632, e-mail: nedko@bgearn.acad.bg
\end{abstract}



Recent discovery of a quantum disordered phase and a spin gap
in the quasi-two-dimensional magnet $CaV_4O_9$ \cite{taniguchi}
has attracted considerable
interest\cite{katoh,ueda,sano,albrecht}. The magnetic system
can be described by the antiferromagnetic
Heisenberg model on a $1/5$-depleted
square lattice containing interacting  plaquette spins and
additional frustrating interactions\cite{ueda}. In the limit of
weak interplaquette and frustrating couplings, the ground state
of the model is a product of resonating
four-spin singlet states formed on each of the elementary plaquettes
of the lattice. In the other limit of strong interplaquette couplings,
the ground state approaches the spin-Peierls phase constructed of
dimers connecting neighboring plaquettes. A number of interesting
questions has arised concerning
the nature of the transition between these two
phases, the possible existence of an intermediate  N\'eel phase,
and the role of geometrical frustration in the formation of
a spin  gap. In this paper we address similar problems  in
plaquette Heisenberg chains, Fig. 1, which are
closely related to the parent two-dimensional depleted
lattice described above. Apart from the pure
 theoretical interest,
we believe that the plaquette spin chains may appear as real objects
in some  magnetic systems. To the best of our knowledge, plaquette
Heisenberg chains have not been  studied except for the
recent work of Katoh and Imada who
 presented a second-order perturbation
estimate for the gap of  the unfrustrated model
in the limit of weak interplaquette interaction\cite{katoh}.
We consider  the simplest Heisenberg plaquette chain
described by the Hamiltonian
\begin{equation}
H= H_1+H_2+H_3\equiv J_1\sum{\bf S_iS_j}+
J_2\sum{\bf S_iS_j}+
J_3\sum{\bf S_iS_j},
\label{ham}
\end{equation}
where the sums run over the dimer $J_1$, plaquette $J_2$, and
diagonal $J_3$ bonds,
respectively, Fig. 1a. For the sake of clarity,
we neglect the second diagonal coupling in
the four-spin plaquettes.
The exchange constants
 $J_1$, $J_2$, and  $J_3$ are supposed to be non-negative. The
so-defined model is perhaps among the simplest ones displaying the
tendency to four-spin plaquette and spin-Peierls dimer ground states in
the limits of weak and strong interplaquette (dimer) interaction $J_1$,
respectively. The inclusion of  $H_3$ in Eq. (1)  opens a
chance to study the role of frustration in the formation of
of a spin gap structure and the relative stability of these two
phases.

Note also that the square of the the local spin operator
 ${\bf S}_{oi}={\bf S}_{ai}+{\bf S}_{bi}$ ( $i=1,2,...,N_p$,
 $N_p=N/4$) is a good quantum number,
so that the "diagonal" plaquette
spin $s_{oi}$ defined by ${\bf S}_{oi}^2=s_{oi}(s_{oi}+1)$
 can be used
to classify the states of the chain by the
set $[s_{o1},s_{o2},..., s_{oN_p}]$ defining a
 sector of states for every given
$s_{oi}:$ $\mid s_{ai}-s_{bi}\mid \leq s_{oi}\leq
s_{ai}+s_{bi}$. Here $s_{ai}$ and $s_{bi}$ are the spins of the
diagonal sites
$a$ and $b$ in the elementary plaquette $i$. Due to the relation
\begin{equation}
s_{oi}(s_{oi}+1)=s_{ai}(s_{ai}+1)+
s_{bi}(s_{bi}+1)+2{\bf S}_{ai}{\bf S}_{bi},
\end{equation}
the local operator ${\bf S}_{ai}{\bf S}_{bi}$ (and the
global one $H_3$), introducing
a geometrical frustration in the plaquette $i$, is diagonal and
explicitly known in every subspace connected to
$[s_{o1},s_{o2},..., s_{oN_p}]$.
In the particular case  $s_{ai}=s_{bi}\equiv s=1/2$,
the number of sectors $2^{N_p}$ corresponds to the number of all
possible sets $[111010..01]$, where $1$($0$) stands for
  $s_{oi}=1$($s_{oi}=0$).

$(i)$ {\it Unfrustrated plaquette chain}, $J_3=0$:

An exact statement concerning the structure of the ground
state of the unfrustrated $J_3=0$
plaquette chain can be deduced from the
Lieb-Mattis theorem\cite{lieb} for
Heisenberg antiferromagnets on bipartite lattices.
 A corollary of the latter
theorem\cite{klein}, valid for the ground state
from the subspace with a given total spin, reads
$\langle {\bf S}_{i}{\bf S}_{j}\rangle >0 $ provided that the pair
of sites
($i,j$) belongs to
one and the same sublattice, and $\langle {\bf S}_{i}{\bf S}_{j}\rangle < 0$
in the opposite case. Applied to the diagonal spins in a given plaquette,
the  sign rule reads $\langle {\bf S}_{ai}{\bf S}_{bi}\rangle > 0$.
 Thus, for a $s=1/2$  plaquette chain one obtains the
  following statement.  {\it The ground state of a $s=1/2$
plaquette chain without frustration is a singlet state and
belongs to the sector $[11...1]$}.
In the general case of  arbitrary site spins, the above sign rule,
combined with Eq. (2), implies
\begin{equation}
s_{oi}(s_{oi}+1)>s_{ai}(s_{ai}+1)+s_{bi}(s_{bi}+1),\hspace{1cm}
i=1,...,N_p.
\end{equation}
 One can check that
for $s_{ai}=s_{bi}=s$ and $s>2$, the diagonal plaquette spin
 $s_{oi}$ has more than one possible
value for every plaquette.

Now, let us consider  unfrustrated $s=1/2$ open chains containing
$N_p$ plaquettes and two end spins of the form shown in Fig. 1a.
We already know that the ground state belongs to the sector $[11...1]$.
There is a simple way to build classes of singlet excited states
using the following construction. Consider the lowest
excited state of the chain
in the sector with exactly one $s_{oi}=0$ plaquette,
say, $[110111]$ for $N_p=6$.
Due to the presence of a singlet dimer bond in one of the
plaquettes, the state
is a  product of the ground states of two independent clusters
of the same form ( containing $N_p=2$ and $N_p=3$ plaquettes
in our example).
If the $s_{oi}=0$ plaquette is situated in the end
of the chain, then the product will contain the state of
a $N_p-1$ plaquette cluster and two singlet dimer states formed
on the $(ab)$ and $(cd)$
end bonds.
This factorization feature reflects the  fact that
 the third spin in a triangle
is completely free provided that the other two   spins
are in a singlet dimer state. We can continue the dividing procedure by
adding new $s_{oi}=0$ plaquettes. The final state, belonging to the
sector $[000000]$, is a spin-Peierls dimer state. Note that the
above classes of states are directly connected to the states of
the periodic chain, namely, the ground state of
a periodic chain with $N_p$
plaquettes in the subspace with one $s_{0i}=0$ plaquette is just the
ground state of an open cluster with $N_p-1$ plaquettes times the
singlet dimer state  on the $i$-th $(ab)$ bond.
 The above construction
is applicable without changes also to the frustrated system.

In Fig. 2 we present exact diagonalization (ED) results
(for periodic chains)
for the correlations of the nearest-neighbor
spins on the plaquette ($ac$)  and dimer ($cd$) bonds.
 It is  seen that already at $J_2=J_1$
the resonating plaquette character  of the short-range order
 is practically  completely
established. In the region near $J_2\approx 0.5J_1$,
the competition
between the dimer and the plaquette type short-range correlations
is strongly pronounced. Notice, however, that
the symmetry of the ground state is not changed for arbitrary
finite $J_2$.

 In Fig. 4 are in particular shown ED data for the spin gap
function $\Delta (J_2)$ of the unfrustrated model
 (open circles).  For small enough
$J_2$, the  numerical data
predict $\Delta \sim J_2^2$, whereas for $J_2\geq 1$ one has
$\Delta \sim J_2$. Here $J_1 \equiv 1$.
In the small $J_2$ limit, the above behavior can also be deduced
from the perturbation theory. Alternatevely, for small enough $J_2$
one can also apply the well-known Dasgupta-Ma eliminating
 procedure \cite{ma}
in order to remove the dimer spins from the Hamiltonian. The
resulting effective model, on the other hand,
 was shown by Schulz \cite{schulz} to be
equivalent to the $S=1$ Heisenberg chain. Thus,
the finite spin-gap structure
 in a $s=1/2$ plaquette chain for small $J_2$
corresponds to a picture where the low-energy physics of
the model is dominated by the dynamics of the "solid" diagonal
spins ${\bf S_{0i}}$ coupled by an effective
antiferromagnetic exchange interaction $J_{eff}\sim J_2^2/J_1$
produced by the spins
on the dimer bonds, i.e., an effective Haldane-type
model with a spin gap
$\Delta \sim J_{eff}exp(-\pi s_0), s_0=1$\cite{affleck}.
Close to the dimer limit $J_2/J_1 \ll 1$,   the ED data
show  spin-spin correlations characteristic of a
spin-one Heisenberg antiferromagnetic chain.

$(ii)$ {\it Frustrated plaquette chain}, $J_3>0$:\\

 Below we proceed to  frustrated periodic chains
containing $N_p$ elementary plaquettes.
It is easy to show that for sufficiently large $J_3$
the dimer eigenstate
\begin{equation}
|\Psi_0 > =\prod_{i=1}^{N_p}\chi_{ab}^{(s)}(i)\chi_{cd}^{(s)}(i)
\end{equation}
is an exact ground state of the model. Here
$\chi_{ab}^{(s)}(i)\equiv
[\alpha_{a}(i)\beta_{b}(i)-\beta_{a}(i)\alpha_{b}(i)]/\sqrt{2}$ and
$\chi_{cd}^{(s)}(i)\equiv
[\alpha_{c}(i-1)\beta_{d}(i)-\beta_{c}(i-1)\alpha_{d}(i)]/\sqrt{2}$,
 $\alpha$ and
$\beta$ being the usual $s^z=\pm 1/2$ states for $s=1/2$.
 This can be achieved most
readily if one  prove that the energy of
the above state
\begin{equation}
E_d(N_p)=-\frac{3}{4}J_1N_p-\frac{3}{4}J_3N_p
\end{equation}
 saturates a lower
bound to the ground state energy for suitable values of the
parameters $J_1$, $J_2$, and $J_3$. To this end, one can use the
simplest cluster decomposition of the
 Hamiltonian $H=\sum_{i=1}^{2N_p}h_i$,
where $h_i=(J_1/2){\bf S_1S_2}+J_2{\bf S_2(S_3+S_4)}+(J_3/2){\bf
S_3S_4}$ is the Hamiltonian of  the cluster shown in Fig. 1b.
The variational principle implies  the following relations for the
ground state energy $E_0(N_p)$
\begin{equation}
E_d(N_p) \geq E_0(N_p) \geq 2N_p \epsilon ,
\end{equation}
where $\epsilon =-(J_1/8)-(J_2/4)+
(J_3/8)-\sqrt{J_1^2-2J_1J_2+9J_2^2}/4$ is
the ground state energy of the cluster. The latter inequality
implies the following  bound above which $|\Psi_0>$ is an
exact ground state
\begin{equation}
J_{3c}=-\frac{1}{2}+\frac{J_2}{2}+\frac{1}{2}\sqrt{1-2J_2+9J_2^2},
\hspace{1cm} J_1 \equiv 1.
\end{equation}
In Fig. 3 we compare this bound with ED data for a cluster
with $N=24$ sites (the
data for $N=16$ are  very close to the presented numerical data).
It is seen a good
theoretical estimate in  view of the fact that we have
used the simplest available cluster.

For the model under consideration one can also use the
following expression for the ground state energy in the
subspace with $n$ diagonal spins in a $s_{0i}=1$ state and
$N_p-n$ diagonal spins in a $s_{0i}=0$ state
\begin{equation}
E^{(n)}(N_p)=E^{(n)}_0(N_p)+J_3n-\frac{3}{4}J_3N_p,
\end{equation}
where $E^{(n)}_0(N_p)$ is the ground state energy of the unfrustrated
Hamiltonian $H_0=H_1+H_2$ in the same subspace. From the
condition $E^{(n)}(N_p)=E_d(N_p)$ one can find the coupling $J_3$
at which the energy levels $E^{(n)}(N_p)$ cross $E_d(N_p)$.
In the case $n=N_p$ one
gets the following bound
\begin{equation}
J_{3c}(N_p)=-\frac{3}{4}-\frac{E_0(N_p)}{N_p},
\hspace{1cm} J_1\equiv 1,
\end{equation}
where $E_0(N_p)$ is the ground state energy of the $J_3=0$ periodic
chain with $N_p$ plaquettes. In Fig. 3
is shown that the latter expression
for a periodic chain with $N_p=6$ plaquettes
exactly represents the numerical ED data for the exact bound above
which the state $|\Psi_0>$ is a ground state of the model. However,
it is not clear if the above expression also describes the
exact absolute bound in the thermodynamic limit\cite{kitatani}.

 In Fig. 4 we  show ED data
for the spin gap function $\Delta(J_2)$ for
$J_3=0.5$. One can clearly see  different
 behavior of $\Delta (J_2)$
before and after the point $J_{2c} \approx 0.53$.
The latter point is close to the estimate
from Fig. 3 at $J_{3c} =0.5$. Notice also that the data in Fig. 4
in the region $J_2 > J_{2c}$  coincide with
 the results in the unfrustrated system. The latter observation
 is a result of Eq. (8) provided that the lowest excited state
 belongs to the sector of the ground state. Finally,
 the ED data show that in the region $J_2 < J_{2c}$
 the $J_3=0.5$ numerical data are in essence independent of $N$
 for  the studied clusters, $N=8$, $16$, and $24$.

$(iii)$ {\it Conclusions}:\\

The above results
demonstrate the role of frustration
in the plaquette chain model. In the frustrated
plaquette chain one finds two different (by symmetry) phases.
 For large enough
$J_2$  and fixed $J_3$, the ground state belongs to the
sector $[11...1]$ and it has features which are
typical of the unfrustrated system.
For small $J_2$ plaquette coupling, the ground state belongs to
the sector $[00...0]$ which is characterized by a spin-Peierls
type dimer state. In principle, one should expect a quantum
first-order phase
transition between the above phases deserving a special consideration.
 In the two-dimensional
analog of the plaquette chain, there is a new role of the frustration.
The latter may also destroy the possible intermediate N\'eel phase
which does not present in the one-dimensional system.


 N.I. was supported by the
National Science Foundation, Grant $\Phi412/94$. J.R.
was supported by DFG, Grant Ri-615/1-2.
The visit of N.B. in the University of Magdeburg was supported by DFG.



\pagebreak

\begin{figure}
\caption{ a) The plaquette chain model studied in the paper. $J_1$,
$J_2$, and $J_3$ are respectively the dimer,  plaquette, and
diagonal antiferromagnetic couplings. The spins leaving on $ab$ and
$cd$ sites in every four-spin plaquette are called diagonal and
dimer spins, respectively.
b) The cluster used to find a lower bound above which the dimer state
is an exact ground state of the plaquette chain.}
\label{fig1}
\end{figure}

\begin{figure}
\caption{ Exact diagonalization data
for the short-range plaquette, $<{\bf S_{a}}(i){\bf S_{c}}(i)>$,
and dimer, $<{\bf S_{c}}(i){\bf S_{d}}(i+1)>$, spin-spin correlators
respectively normalized by the factors $-1/2$ and $-3/4$,
vs  $J_2$. $J_1 \equiv 1$. }
\label{fig2}
\end{figure}

\begin{figure}
\caption{The bounds $J_{3c}$ vs $J_2$ above which the state
$|\Psi_0>$, Eq. (4), is an exact ground state of the frustrated
plaquette chain. The solid line represents the
 analytic result, Eq. (9), for $N_p=6$. The dashed line is the
 lower bound, Eq. (7),
obtained with the cluster shown in Fig. 1b. The open circles are
numerical ED results for a periodic chain with $N_p=6$
plaquettes, i.e., $N=24$ sites. $J_1 \equiv 1$.}
\label{fig3}
\end{figure}

\begin{figure}
\caption{Exact diagonalization results for the
spin gap function $\Delta (J_2)$ of the $J_3 =0$
and $J_3=0.5$ plaquette chain. The extrapolation is
deduced from the $N=8$, $16$, and $24$ site clusters.
$J_1 \equiv 1$.}
\label{fig4}
\end{figure}
\end{document}